\documentclass{llncs}
\usepackage{graphicx}
\usepackage{adjustbox}
\usepackage{xcolor}
\usepackage{hyperref}
\usepackage{url}

\newcommand{\code}[1]{\texttt{\small{}#1}}

\begin{document}
\title{Provenance-Centered Dataset of Drug-Drug Interactions}
\titlerunning{Dataset of Drug-Drug Interactions}  
%
\author{Juan M. Banda\inst{1} \and Tobias Kuhn\inst{2,3} \and
Nigam H. Shah\inst{1} \and Michel Dumontier\inst{1}}
\authorrunning{Juan M. Banda et al.} 
%
\tocauthor{Juan M. Banda, Tobias Kuhn, Nigam H. Shah, Michel Dumontier}
\institute{Stanford University - Center for Biomedical Informatics Research, 1265 Welch Road, Stanford, CA, 94305, USA\\
\email{jmbanda@stanford.edu}, \email{nigam@stanford.edu}, \email{michel.dumontier@stanford.edu}
\and
Department of Humanities, Social and Political Sciences, ETH Zurich, Switzerland\\
\and
Department of Computer Science, VU University Amsterdam, Netherlands\\
\email{tokuhn@ethz.ch}}

\maketitle              

\begin{abstract}
Over the years several studies have demonstrated the ability to identify potential drug-drug interactions via data mining from the literature (MEDLINE), electronic health records, public databases (Drugbank), etc. While each one of these approaches is properly statistically validated, they do not take into consideration the overlap between them as one of their decision making variables. In this paper we present LInked Drug-Drug Interactions (LIDDI), a public nanopublication-based RDF dataset with trusty URIs that encompasses some of the most cited prediction methods and sources to provide researchers a resource for leveraging the work of others into their prediction methods. As one of the main issues to overcome the usage of external resources is their mappings between drug names and identifiers used, we also provide the set of mappings we curated to be able to compare the multiple sources we aggregate in our dataset.
\keywords{drug-drug interactions, nanopublications, data mining}
\end{abstract}

\section{Introduction}
Studies analyzing costs over time have shown that adverse drug reactions (ADRs) cost over \$136 billion a year \cite{REF1}. One significant cause of ADRs are drug-drug interactions (DDIs) which greatly affect older adults due to the multiple drugs they are taking \cite{REF4}. A DDI occurs when the effect of any given drug is altered by another drug which results in an unpredictable effect. While new drugs, before market approval, are tested in both \textit{in vivo} and \textit{in vitro} methods \cite{REF6}, it is unfeasible to test their interactions with all other approved and experimental drugs. In the recent years, computational approaches have been trying to infer potential DDI signals using a wide variety of sources \cite{REFINDI,REFSRINI,REFTWOSIDES,REFSANTI}, however these methods produce thousands of statistically plausible predictions \cite{REFINDI,REF24}, thus making the task of testing them in an experimental setting impractical.

We are therefore facing the need of combining data from various disparate sources, but at the same time have to remain aware of their provenance due to differences in individual quality and overall confidence. Depending on their primary source and their provenance path, two data entries about the same DDI might substantially increase our confidence about a relation. By relying on Semantic Web technologies such as RDF -- and more specifically nanopublications \cite{groth2010isu} -- these provenance paths can be represented in an explicit and uniform manner. In this paper we introduce LInked Drug-Drug Interactions (LIDDI), a dataset that consolidates multiple data sources in one cohesive linked place that allows researchers to have immediate access to multiple collections of DDI predictions from public databases and reports, biomedical literature, and methods that take different and sometimes more comprehensive approaches. 

\begin{figure}[tp]
\centering
    \includegraphics[width=1.0\textwidth]{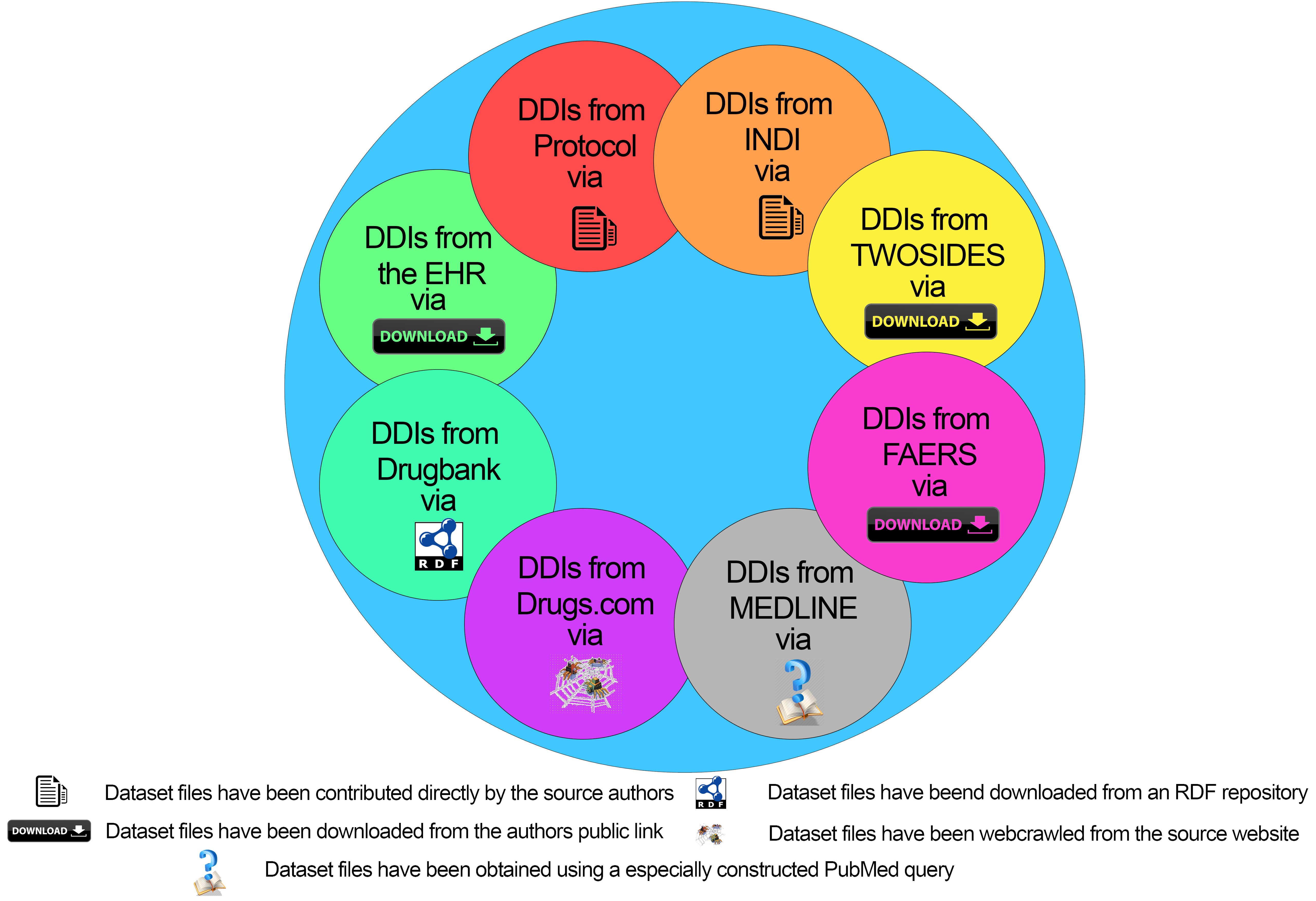}
\caption{Overview of the Linked Drug-Drug Interactions dataset and the data sources that it incorporates and their original formats}
\label{fig:datasetflow}
\end{figure}

With great potential for use in drug safety studies, LIDDI provides the linking components to branch from DDIs into individual properties of each drug via Drugbank and UMLS, as well as mappings to other types of biomedical ontologies which are used to describe drug-drug interactions. In terms of direct applications, the resources in our dataset allow researchers to quickly determine if there is support by other DDI sources for their own predictions, allowing them to evaluate them via consensus rather than only by independent statistical measures (AUROC values, odds ratios, etc). LIDDI offers all the mappings needed to bridge the incorporated resources into a single comparable entity, providing extra value for researchers looking to bridge the data sources we have connected, for example Drugbank drug identifiers to UMLS Concept Unique Identifiers (CUI) to Medical Subject Headings (MeSH) codes, opening endless possibilities for reuse.

To our knowledge, this is the first public dataset exemplifying the use of nanopublications for the integration of knowledge from multiple diverse data sources in the field of drug safety surveillance. We believe it will facilitate cross-referencing of results between different domains not previously available, as well as the enrichment of drugs involved in the DDIs thanks to the linkages provided.

\section{Methods}
\subsection{Nanopublications with Trusty URIs}

Nanopublications \cite{groth2010isu} are a concept to use Semantic Web techniques (most importantly RDF and named graphs) to closely link data to their provenance and meta-data in a uniform manner. The vision is that small data packages should become the primary format for research outputs instead of narrative articles \cite{mons2011naturegen}. Technically, a nanopublication consists of an assertion graph with triples expressing an atomic statement (about drug-drug interactions in our case), a provenance graph that reports how this assertion came about (e.g. where it was extracted from or what mechanism was used to derive it), and a publication information graph that provides meta-data for the nanopublication (such as its creators and a timestamp).

Trusty URIs \cite{kuhn2014eswc,kuhn2015tkde} are a recent proposal to make URI links for digital artifacts verifiable, immutable, and permanent. Such URI identifiers contain a cryptographic hash value calculated on the content of the represented digital artifact --- i.e. RDF content of nanopublications in our case --- in a format-independent way. With trusty URIs, any reference to an artifact thereby comes with the possibility to verify with 100\% confidence that a retrieved file really represents the correct and original state of that resource.

\subsection{Data Schema}
\begin{figure}[tp]
\centering
    \includegraphics[width=\textwidth]{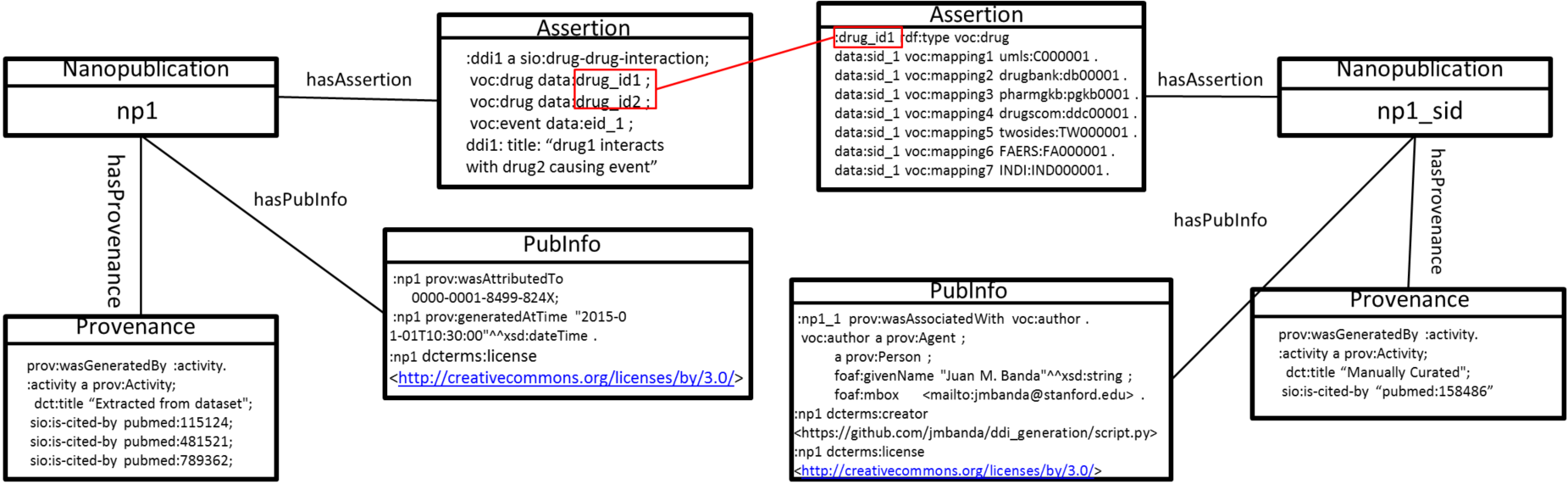}
\caption{Dataset Schema shows two types of nanopublications included in the dataset. The first captures the DDI, and the second captures drug mappings.A third one not show captures the event mappings.}
\label{fig:datasetschema}
\end{figure}

 We represent DDIs in the following way: We instantiate the DDI class from the Semantic science Integrated Ontology (SIO) \cite{dumontier2014biomedsem}. We define a vocabulary in our own namespace, as is normally done with Bio2RDF \cite{callahan2013eswc}, to assign domain-specific predicates to point from the DDI to each drug, and to the clinically-relevant event resulting from the drug interaction. This assertion is stored in the assertion graph of the nanopublication. We also create nanopublications to capture the mappings between each drug and our seven sources. We use BioPortal PURL identifiers for UMLS URIs. We use the PROV ontology \cite{lebo2013prov} to capture the provenance of each DDI by linking the assertion graph URI to a PROV activity. This activity is linked to the software used to generate the drug and event mappings, as well as to the citation for the method, the generation time, and the dataset where the DDIs were extracted from. In this last case we also provide a direct link to the original dataset used. Figure \ref{fig:datasetschema} highlights our general schema with a particular example for a DDI that contains a drug-drug set. 

\section{Data sources aggregated}

In this section we provide details about the transformations we performed on each of the data sources to be able to normalize them and use them together. Due to space constraints we provide a very high-level view of how each source procures and generates its data as this lies out of the scope of this dataset paper. All the mappings provided within our dataset for UMLS concept unique identifiers (CUI), RxNORM codes, MeSH codes and Drugbank identifiers, were done as an initial step once we acquired all our sources. 

\paragraph{Electronic Health Records (EHR).}
In \cite{REFSRINI} the authors used Stanford Translational Research Integrated Database Environment (STRIDE) dataset comprising 9 million unstructured clinical notes corresponding to 1 million patients that span a period of 18 years (1994--2011) to mine DDIs from the EHR. Published as a comma delimited file in the supplemental materials, we use the highest confidence predictions as the initial basis of our dataset in terms of drugs (345) and events (10). This dataset provides drug-drug-event sets of UMLS concept unique identifiers (CUI), drug/event names in string literals.

\paragraph{FDA Adverse Event Reporting System (FAERS).}
We analyzed over 3.2 million reports found on the FDA Adverse Event Reporting System (FAERS), since there are no established algorithmic ways of determining the statistical significance of a drug-drug interaction from FAERS reports without any additional external information, we set a threshold of each DDI to have at least ten appearances in different FAERS reports for it to appear on our dataset. In this data source, the drugs are given in string literals that normalized to RxNorm drug names and the events normalized to MeDRA. 

\begin{figure}[tp]
\centering
    \includegraphics[width=0.95\textwidth]{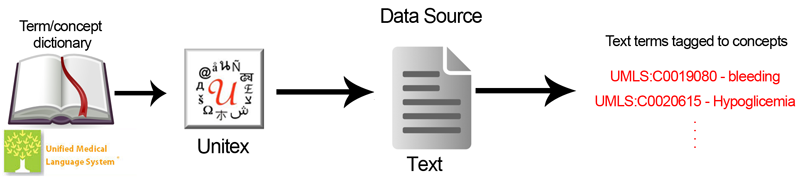}
\caption{Text-tagging pipeline. We build a term/concept dictionary from ontologies in UMLS. We then proceed to tag concepts found in any piece of text from any source via Unitex, resulting in a bag of UMLS CUIs for each input.}
\label{fig:annotation}
\end{figure}

\paragraph{DrugBank.}
We used the RDF versions of Drugbank \cite{REFDRUGBANK} made publicly available through Bio2RDF. The used version was dated as 2013-07-25. We proceeded to strip the Drugbank identifiers and their labels into string literals mapped into RxNorm/SNOMED string literals. In order to extract DDIs we used the \code{drugbank\_vocabulary:Drug-Drug-Interaction} type to match interacting drugs, we then used the \code{rdfs:label} to extracted the event by concept tagging the indication section with a modified version of the NCBO annotator used in \cite{REFSRINI}. This concept tagging process is described on Figure ~\ref{fig:annotation}. 

\paragraph{Drugs.com.}
In order to mine DDIs from this resource, we used a web-crawling agent developed in Python. We programmatically checked the individual drugs.com page for each of the 345 drugs in our set and extracted all their interacting drugs. For each of the interacting drugs we build the event part of our DDI and by annotating the complementary free-text section, the same way as on Figure ~\ref{fig:annotation}, which describes the interaction (similar to the Drugbank indication).

\paragraph{MEDLINE.}
In order to find potential DDIs signals in the biomedical literature, we have adapted a clever method developed by Avilach et al. \cite{avillachDDI}, that allows for the use of a query with certain MeSH terms and qualifiers to find adverse events in the literature. We modified this query by adding a MeSH term for "Drug interactions" and modified the search to allow for two drugs instead of one from the original approach. These modifications of the method with some manipulation produces drug-drug-event sets.

\paragraph{TWOSIDES.}
Provided as a download by Tatonetti et al. \cite{REFTWOSIDES}, TWOSIDES contains polypharmacy side effects for pairs of drugs. Mined from FDA’s FAERS, this methods is designed to complement modern signal detection approaches, providing stratification, dampening or removing the effect of co-variates, without needing to divide drug-exposed reports into strata. This resource is available in a CSV format and features drugs in RxNorm normalized string literal form and events in UMLS CUI form.

\paragraph{INferring Drug Interactions (INDI).}
Currently available on a website for checking drug-drug interactions and provided to us by the authors \cite{REFINDI} in a comma separated file. This prediction method infers both pharmacodynamic and pharmacokinetic interactions based on their similarity to existing known adverse events that result from a common metabolizing enzyme (CYP). The provided resources consisted of drug-drug sets (no events) in UMLS CUI form.

\paragraph{Similarity-based Modeling Protocol (Protocol).}
Provided to us by the authors, Vilar et al. \cite{REFSANTI}, this protocol integrates a gold standard of DDIs with drug similarity information extracted from sources like: 2D and 3D molecular structure, interaction profile, target similarities, and side-effect similarities. This method generates drug interaction candidates that are traceable to pharmacological or clinical effects. We were provided with the resulting drug-drug (no events) sets with Drugbank identifiers.  

\section{Statistics and Access}


Table \ref{tab:table1} shows the total number of DDIs found with support in each of the multiple sources we integrated. Note that these are not unique DDIs in the sense that we have a drug1-drug2-event set that is equivalent to drug2-drug1-event set as the directionality does not matter, thus counting them twice. We left this reflexive drug drug interactions in our dataset to conform to the standard employed by bio2RDF's DrugBank repository, in order to not limit the discovery of any potential interaction when a query is performed on any given single drug.

\begin{table}[tp]
\begin{center}
\caption{\label{tab:table1}LIDDI DDI event distribution between sources}
\resizebox{0.99\textwidth}{!}{
\begin{tabular}{ | c | c | c | c | c |c | c | c | c | }
\hline
\textbf{Event Name} &	\textbf{EHR} &	\textbf{MEDLINE} &	\textbf{Drugbank} &	\textbf{Drugs.com} &	\textbf{FAERS} &	\textbf{TWOSIDES} &	\textbf{INDI* } &	\textbf{Protocol* } \\
\hline
Arrhythmia &	700 &	68 &	286 &	3,148 &	100 &	4,632 &	NA* &	NA* \\
Bradycardia &	254 &	88 &	408 &	4,896 &	194 &	4,824 &	NA* &	NA* \\
Hyperkalaemia &	1,888 &	42 &	422 &	4,248 &	146 &	3,840 &	NA* &	NA* \\ 
Hypoglycaemia &	1,460 &	386 &	796 &	6,214 &	104 &	5,150 &	NA* &	NA* \\
Long QT syndrome &	14 &	270 &	334 &	3,510 &	2 &	0 &	NA* &	NA* \\ 
Neutropenia &	4,608 &	192 &	402 &	4,218 &	616 &	3,702 &	NA* &	NA* \\
Pancytopenia &	1,880 &	4 &	270 &	3,146 &	148 &	5,440 &	NA* &	NA* \\
Parkinsonism &	144 &	0 &	566	& 5,978 &	70 &	884 &	NA* &	NA* \\
Rhabdomyolysis &	122 &	198 &	392 &	3,842 &	214 &	3,264 &	NA* &	NA* \\
Serotonin syndrome &	896 &	0 &	384 &	3,960 &	122 &	1,094 &	NA* &	NA* \\
\hline
Total: &  11,966  & 1,248 & 4,260  & 43,160 & 1,716 & 32,830 & 8,370 & 224 \\
\hline
\end{tabular}
}
\end{center}

\end{table}

The dataset in its entirety contains a total of 98,085 nanopublications out of which 345 are used for drug mappings, 10 for event mappings and the remainder for DDIs extracted from the data sources we used. The dataset has a total of 392,340 graphs (four per nanopublications) and 2,051,959 triples, taking 723 MB in nquads representation.

\setcounter{footnote}{0}

LIDDI can be accessed as a bulk download via figShare \cite{figshare} and SPARQL endpoint at: \url{http://http://liddi.stanford.edu:8890/sparql}.The dataset is also made available via the recently installed nanopublication server network \cite{kuhn2015iswc}. The command `\texttt{np}' from the Java nanopublication library\footnote{\url{https://github.com/Nanopublication/nanopub-java}} can be used to download the dataset \cite{nanopubindex2015liddi}:
\begin{quote}\small\tt
\$ np get -c -o out.trig RA7SuQ0e661LJdKpt5EOS2DKykf1ht9LFmNaZtFSDMrXg
\end{quote}

\section{Potential for applications and future work}

The first application that this dataset has been used for involves ranking DDI predictions from EHR data. By taking the EHR data source published by \cite{REFSRINI} and using this resource, researchers are ranking predictions in order to determine which ones are more feasible for experimental evaluation. By leveraging such diverse resources, the researchers have been able to use the data sources as voting mechanisms for certain DDIs to be prioritized over others. Similar studies can be performed to determine the priority of the other data sources and their predictions just using LIDDI and selecting a different resource to evaluate. The authors of each data source could enhance their own predictions by using their method and consider other sources we compiled in this dataset as gold-standards or at least 'silver-standards' as there is no de facto community-wide gold standard for drug safety when it comes to drug-drug interactions. Multiple applications can revolve around enhancing the DDIs listed here by enriching the drugs and events with properties available in the Drugbank, RxNorm and MeSH linked graphs that are not easily accessible via traditional means. This might lead to even further linkage of other resources tying all things back to a described DDI and helping to further explain its reason and impact.
As more researchers become aware of this resource they should be encouraged to contribute to it by adding their own DDI predictions, thus enhancing the overall availability of DDI predictions sets for scientific comparisons. We will continue to contact authors and map their data sets into LIDDI to have a more diverse and rich set of resources. We also plan at some point incorporate more drugs and more events into the resource as they become available for it to have a wider coverage of the current drug space.

\bibliographystyle{abbrv}
\bibliography{main}

\end{document}